\documentstyle[12pt,axodraw,epsfig]{article}
\def\ltsim{\lower3pt\hbox{$\, \buildrel < \over \sim \, $}}
\def\gtsim{\lower3pt\hbox{$\, \buildrel > \over \sim \, $}}
\def\be{\begin{equation}}
\def\ee{\end{equation}}
\def\ba{\begin{eqnarray}}
\def\ea{\end{eqnarray}}
\def\ga{\mathrel{\raise.3ex\hbox{$>$\kern-.75em\lower1ex\hbox{$\sim$}}}}
\def\la{\mathrel{\raise.3ex\hbox{$<$\kern-.75em\lower1ex\hbox{$\sim$}}}}
\newcommand{\sect}[1]{\section{#1}\setcounter{equation}{0}}

\newcommand{\bi}[1]{\bibitem{#1}}

\begin{document}
\baselineskip=16pt
\begin{titlepage}
\rightline{OUTP-00-12P}
\rightline{hep-ph/0003207}
\rightline{March 2000}  
\begin{center}

\vspace{0.5cm}

\large {\bf  $''+-+''$  Brane Model Phenomenology}
\vspace*{5mm}
\normalsize

{\bf Stavros
Mouslopoulos\footnote{s.mouslopoulos@physics.ox.ac.uk}} and {\bf Antonios
Papazoglou\footnote{a.papazoglou@physics.ox.ac.uk}}

\smallskip 
\medskip 
{\it Theoretical Physics, Department of Physics, Oxford University}\\
{\it 1 Keble Road, Oxford, OX1 3NP,  UK}
\smallskip

\vskip0.6in \end{center}
 
\centerline{\large\bf Abstract}

We explore the phenomenology of the recently proposed $''+-+''$ brane
model which has a characteristic anomalously light first Kaluza-Klein mode. We
consider the  processes $e^+e^-\rightarrow \mu^+\mu^-$ and the
Kaluza-Klein production $e^+e^- \rightarrow \gamma {\rm KK}$ giving
missing visible energy. These in combination with the latest
Cavendish experiments place severe bounds on the parameter space of the model. We also discuss how forthcoming
experiments can test the model for  ``natural'' range of the parameters.  

\vspace*{2mm} 

\end{titlepage}

\sect{Introduction}

There has been a lot of interest during the past two years in models where
the Standard Model (SM) fields are localized on a 3-brane in a higher dimensional
spacetime \footnote{This idea actually dates back to the early eighties \cite{80s}}. The motivation behind these constructions was the
description and even solution of the long standing Planck hierarchy
problem by geometrical means. The striking feature of these models is
that they make dramatical phenomenological
predictions which can be directly confronted with current and future
collider experiments. 

We can generally divide the recent  brane-models into two classes. The first one
pioneered by Antoniadis, Arkani-Hamed, Dimopoulos and Dvali
\cite{large}, assumes factorizable geometry along the extra
dimensions. The Planck hierarchy is then explained by the largeness of
the compactification volume that suppresses the Planck scale and gives
rise to a TeV fundamental mass scale ($M_{\rm Pl}^2=M^{n+2}V_n$). The
size of the $n$ new  dimensions  can be
as large as a millimeter without coming in conflict with
experiment.  The second class of models which were  considered by
Randall and Sundrum (RS)
\cite{RS} assumes non-factorizable geometry (essentially spacetime is  a slice of $AdS_5$) that associates each
position on the extra dimension with different length scales, and thus
different energy ones. The original RS construction consists of two
parallel 3-branes of opposite tension sitting
on the fixed points of an $S^1/Z_2$ orbifold. An exponential 
``warp''
factor in the metric then generates a ratio of the mass scales between the 
two branes
that could be $\mathcal{O}$$\left(10^{15}\right)$ although the size of 
the
orbifold is of the order of Planck length. Assuming that the
fundamental mass scale on the positive brane is of the order of
$M_{\rm Pl}$ we can readily get a mass scale on the negative brane of the order
the electroweak scale, thus solving the Planck hierarchy problem. In this
model the 
compactification radius need only be some 35 times larger than the Planck 
length. However, living on a negative tension brane is generically
problematic as far as cosmology is concerned  unless we impose
non-trivial constrains on the parameters of the model (see \cite{cosm} for
complete discussion).

\begin{figure}[t]
\begin{center}
\begin{picture}(300,100)(0,50)

\BCirc(150,100){50}
\DashLine(100,100)(200,100){3}

\GCirc(100,100){5}{0.6}
\GCirc(200,100){5}{0.6}

\GCirc(183,136){5}{0}
\GCirc(183,64){5}{0}

\Text(85,100)[]{$+$}
\Text(215,100)[]{$+$}
\Text(190,145)[]{$-$}
\Text(190,55)[]{$-$}
\Text(172,135)[]{$L_1$}
\Text(190,110)[]{$L_2$}

\LongArrowArc(150,100)(58,4,45)
\LongArrowArcn(150,100)(58,45,4)
\Text(209,127)[]{$x$}

\SetWidth{2}
\LongArrow(150,100)(150,115)
\LongArrow(150,100)(150,85)

\end{picture}
\caption{$''+-+''$ model with two $''+''$ branes at the fixed points
of a $S^1/Z_2$ orbifold and
a $''-''$ brane in between.}\label{+-+}
\end{center}
\end{figure}
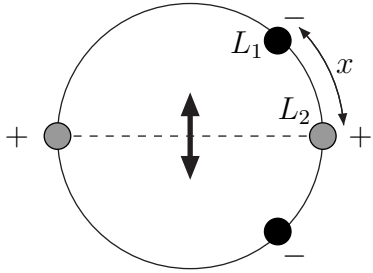

In view of this difficulty Kogan, Mouslopoulos, Papazoglou, Ross and
Santiago \cite{mill} proposed a brane configuration with two $''+''$
branes sitting on the fixed points of an $S^1/Z_2$ orbifold and an
intermediate 
$''-''$ brane (see 
Fig.~\ref{+-+}). In this $''+-+''$ model our universe is the third
$''+''$ brane and a desired hierarchy can be generated by adjusting
the size of the orbifold and the position of the intermediate $''-''$
brane. The above contruction appears to be special because the
equivalent quantum mechanical problem, that arises from the Einstein
equations after some coordinate and field redefinitions, has a double ``volcano'' form
and supports in addition to the bound state graviton zero mode, an anomalously light
and strongly coupled first KK mode ``bound'' state. In
Ref. \cite{mill} we
presented a preliminary study of the $e^+e^- \rightarrow \mu^+\mu^-$
cross section for a specific value of the $w$ and $k$ parameters (see
section 2 for definitions) paying particular regard to the
characteristic behaviour of the first KK mode.

It is worth mentioning that in the context of \cite{mill} if one drops
the requirement of solving the hierarchy problem, one finds the exotic
possibility of ``bigravity''. In this case,  gravitational attraction  is a result of the exchange of
ordinary 4D graviton plus an ultralight first KK state with Compton wavelength
which might be  $1\%$ of the size of the observable
universe, thus generating modifications of gravity not only at small
but also at ultra-large distances. The latter characteristic was also
found  in a brane construction by Gregory, Rubakov and Sibiryakov
\cite{Rub}, which was recently shown to be closely related to the
$''+-+''$ model (see \cite{Kog}). However, as it is widely known \cite{pilo} the
modulus corresponding to the moving negative tension brane in both cases is
a physical ghost, something very unattractive for these 
models. Nevertheless, as proposed by \cite{multi} the
stabilization mechanism might offer a way out of this problem.

In the  present paper we will not discuss the ``bigravity''
possibility, but concentrate on the case that the $''+-+''$ model can offer a
resolution to the  Planck hierarchy problem. Our aim is to further
develop its  collider
phenomenology for different values of the $w$ and $k$ parameters and to
include the constraints from missing
energy processes. The region of the parameter space that can be
considered as ``natural'', in the sense that they solve the hierarchy
problem, is examined in detail. It turns out that different parts  of the parameter
space are sensitive to different type of processes, thus giving us a
clear picture of the expected signatures of the model.

The organization of the paper is as follows. In the next Section we
review the main characteristics of the $''+-+''$ model.  In Section 3 we discuss
the phenomenology of the $e^+e^- \rightarrow \mu^+\mu^-$ and the
$e^+e^- \rightarrow \gamma {\rm KK}$ processes for different values of
the warp factor $w$ and $k$. Finally, we present our conclusions.

\sect{Characteristics of the $''+-+''$ model}

The $''+-+''$ model presented in \cite{mill} (see 
Fig.~\ref{+-+}) has as a starting point the action:
\be
S=\int d^4 x \int_{-L_2}^{L_2} dy \sqrt{-G} 
\{-\Lambda + 2 M^3 R\}-\sum_{i}\int_{y=L_i}d^4xV_i\sqrt{-\hat{G}^{(i)}}
\ee
where $M$ is the 5-D fundamental scale, $\Lambda$ is the bulk cosmological constant,  $\hat{G}^{(i)}_{\mu\nu}$ is the induced metric on the branes
and $V_i$ their tensions. The $''+''$ branes are situated at the fixed
points $L_0=0$, $L_2$ and the $''-''$ brane at $L_1$.

Taking the  metric ansatz that respects 4D Poincar\'{e} invariance to
be \footnote{Here we ignore the radion field which can be used for
modulus stabilization. For details see \cite{rad}.}:
\be
ds^2=e^{-2\sigma(y)}\eta_{\mu\nu}dx^\mu dx^\nu +dy^2
\ee
the Einstein equations give a solution for the $\sigma(y)$ function:
\be
\sigma(y)=k\left\{L_1-\left||y|-L_1\right|\right\}
\ee
with the requirement that the brane tensions are tuned to $V_0=-\Lambda/k>0$,
$V_1=\Lambda/k<0$, \mbox{$V_2=-\Lambda/k>0$}. The parameter $k$ is  a
measure of the curvature of the bulk and is given by
$k=\sqrt{\frac{-\Lambda}{24M^3}}$. The 4D Planck mass, the fundamental
scale $M$ and the parameter $k$ are related by the equation:
\be
M_{\rm Pl}^2=\frac{M^3}{k}\left[1-2e^{-2kL_1}+e^{-2k(2L_1-L2)}\right]
\label{Planck}
\ee

It is obvious that for large enough $kL_1$ and  $k\left(2L_1-
L_2\right)$ the
three mass scales $M_{\rm Pl}$, $M$, $k$ can be taken to be of the same
order.  The ``warp'' factor responsible for the generation of the desired
hierarchy on the third brane is simply:
\be
w=e^{-\sigma\left(L_2\right)}=e^{-k\left(2L_1-L_2\right)}
\ee

The KK spectrum is determined  by considering the (linear)
fluctuations of the metric ansatz. It turns out that there is always
a zero mode that corresponds to the ordinary 4-D graviton. The KK
masses are determined by the the quantization determinant:
\be
\renewcommand{\arraystretch}{1.5}
\left|\begin{array}{cccc}J_1\left(\frac{m}{k}\right)&Y_1\left(\frac{m}{k}\right)
&\phantom{-}0&\phantom{-}0\\0&0&\phantom{-
}J_1\left(\frac{m}{k}e^{k(2L_1-L_2)}\right)&\phantom{-
}Y_1\left(\frac{m}{k}e^{k(2L_1-L_2)}\right)\\J_1\left(\frac{m}{k}e^{kL_1}\right)&Y_1\left(
\frac{m}{k}e^{kL_1}\right)&\phantom{-}J_1\left(\frac{m}{k}e^{kL_1}\right)&\phantom{-
}Y_1\left(\frac{m}{k}e^{kL_1}\right)\\J_2\left(\frac{m}{k}e^{kL_1}\right)&Y_2\left(
\frac{m}{k}e^{kL_1}\right)&-J_2\left(\frac{m}{k}e^{kL_1}\right)&-
Y_2\left(\frac{m}{k}e^{kL_1}\right)\end{array}\right|=0
\ee
(where we have suppressed the subscript $n$ on the masses $m_n$)

Denoting the separation between the last two branes by  $x=k(L_2-L_1)$
we calculate the mass of the first KK state:
\be
m_1=2kwe^{-2x}
\label{m1}
\ee

Numerically, we find that the masses of the remaining KK states depend in a
different way on the parameter $x$. The masses of the second state and the
spacing $\Delta m$ between the subsequent states have the 
form:
\ba
m_2&\approx&4kwe^{-x}\\
\Delta m&\approx&\varepsilon kwe^{-x}
\label{m2}
\ea
with $\varepsilon$ a number between 1 and 2. The spacing only approaches a 
constant for high enough
levels when the arguments of the Bessel functions become much greater than one.

The interaction of the KK states to the SM particles is found as in
Ref. \cite{mill} to be:
\be
{\mathcal{L}}_{int}=-\frac{1}{M_{\rm Pl}}h_{\mu\nu}^{(0)}(x)T_{\mu\nu}\left(x\right)-
\sum_{n>0}\frac{1}{c_n}h_{\mu\nu}^{(n)}(x)T_{\mu\nu}\left(x\right)
\ee
where $T_{\mu\nu}$ is the stress energy momentum tensor of the SM
Lagrangian, $h_{\mu\nu}^{(0)}(x)$ the 4-D graviton,
$h_{\mu\nu}^{(n)}(x)$ the $n$th level KK state and $c_n$ the
corresponding coupling
suppressions of them.

The coupling suppression of the first KK mode is 
independent of $x$ and  equal to 
\be
c_1=wM_{\rm Pl}
\label{c1}
\ee
while the coupling suppressions  of the higher modes are enhanced relative 
to the lowest  mode by a factor proportional to $e^x$.

The above masses and coupling suppressions have been computed in the
region of large $x$  were the Bessel functions can be approximated by the first terms
of their power series (note that since the parameter $x$ appears in
exponentials we actually have very good approximations for $x\ga
3$). However for lower $x$'s the above approximations break down and the first mode is
not so different from the other KK states. In the extreme case where the
positions of the second and the third brane coincide, the positive
brane disappears and we obtain the
original RS model. The phenomenology of the latter
has been extensively studied in  \cite{RSph}.

\newpage

\section{Phenomenology}

In this Section we will present a discussion of the phenomenology of the 
KK modes to be expected in high energy colliders, concentrating on the simple and
sensitive to new physics  processes 
$e^+e^-\rightarrow\mu^+\mu^-$ (this analysis is readily generalized to include 
$q\bar{q}$, $gg$ initial and final states) and $e^+e^-\rightarrow\gamma+
\mbox{\textit{missing energy}}$. Since the characteristics of the
phenomenology depend on the parameters of the model ($w$,$k$,$x$) we
explore the
regions of the parameter space that are of special interest (\textit{i.e.} give
hierarchy factor $\mathcal{O}$$\left(10^{15}\right)$ and do not
introduce a new hierarchy between $k$ and $M$ as seen from equation (\ref{Planck})).  

\subsection{ $e^+e^-\rightarrow\mu^+\mu^-$ process}

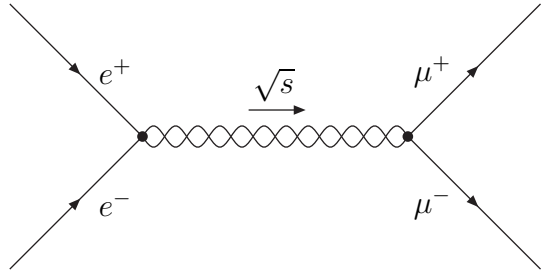
\begin{figure}[b]
\begin{center}
\begin{picture}(300,100)(0,50)
\ArrowLine(50,50)(100,100)
\ArrowLine(50,150)(100,100)
\ArrowLine(200,100)(250,150)
\ArrowLine(200,100)(250,50)
\Vertex(100,100){2}
\Vertex(200,100){2}
\LongArrow(140,110)(160,110)
\Photon(100,100)(200,100){4}{6}
\Photon(100,100)(200,100){-4}{6}
\Text(90,125)[]{$e^+$}
\Text(90,75)[]{$e^-$}
\Text(210,125)[]{$\mu^+$}
\Text(210,75)[]{$\mu^-$}
\Text(150,120)[]{$\sqrt{s}$}
\end{picture}
\caption{$e^+e^-\rightarrow \mu^+ \mu^-$}\label{feyn:diag}
\end{center}
\end{figure}

Using the Feynman rules of Ref. \cite{Lyk1} the contribution of the KK modes to 
$e^+e^-\rightarrow\mu^+\mu^-$ is given by
\be
\sigma\left(e^+e^-\rightarrow\mu^+\mu^-\right)=\frac{s^3}{1280\pi}|D(s)|^2
\ee
where $D(s)$ is the sum over the propagators multiplied by the
appropriate coupling suppressions:
\be
D(s)=\sum_{n>0}\frac{1/c_n^2}{s-m_n^2+i\Gamma_n m_n}
\ee
and $s$ is the center of mass energy of $e^+e^-$.

Note that the bad high energy behaviour (a violation of perturbative unitarity) of this cross section is expected since 
we are working with an effective - low
energy non-renormalizable theory of gravity. We assume our  effective theory is valid up to 
an energy scale
$M_{s}$ (which is ${\mathcal{O}}({\rm TeV})$), which acts as an ultraviolet 
cutoff. The theory that applies  above this scale
is supposed to give a consistent description of quantum gravity. Since this is 
unknown we are only able to determine the contributions of the KK states with 
masses less than this
scale. This means that the summation in the previous formula should
stop at the KK mode with mass near the cutoff.

For the details of the calculation it will be important to know the
decay rates of the KK states. These are given by:
\be
\Gamma_n=\beta\frac{m_n^3}{c_n^2}
\label{gam}
\ee
where $\beta$ is a dimensionless constant that is between
$\frac{39}{320\pi}\approx0.039$ (in the case that the KK is light
enough, \textit{i.e.} smaller than $0.5 {\rm MeV}$, so that it  decays only to 
massless 
gauge bosons and neutrinos) and $\frac{71}{240\pi}\approx0.094$ (in the case where
the KK is heavy enough that can decay to all SM particles).

If we consider $w$ and $k$ fixed, then when
$x$ is smaller than a certain value $x_0=x_0(w,k)$ we have a widely spaced discrete spectrum (from
the point of view of TeV physics) close to the one of the
RS case with cross section at a KK resonances of the form
$\sigma_{res} \sim s^3/m_ n^8$ (see \cite{RSph}).       
For the discrete 
spectrum there is always a range of values of the $x$ parameter so that the
KK resonances are in the range of energies of
collider experiments. 
In these cases we calculate the excess over the SM contribution which would have been seen either by direct
scanning if the resonance is near the energy at which the experiments
actually run or by means of the process
$e^+e^-\rightarrow\gamma\mu^+\mu^-$ which scans a continuum of energies 
below the center of mass energy of the experiment 
(of course if $k$ 
is raised the KK modes become heavier and there will be a value for which the 
lightest KK mode is above the experimental limits). 

For values of $x$ greater than $x_0(k,w)$ the spacing in the spectrum is
so small that we can safely consider it to be continuous. At this
point we have to note that we consider that the ``continuum'' starts at the point
where the convoluted KK resonances start to overlap. In this case
we substitute in $D(s)$ the sum for $n\geq2$ by an integral over the
mass of the KK excitations, \textit{i.e.}
\be
D(s)_{KK}\approx\frac{1/c_1^2}{s-m_1^2+i\Gamma_1 m_1}+
\frac{1}{\Delta m\; c^2} \int_{m_2}^{M_s}dm\; \frac{1}{s-m^2+i\epsilon}
\ee
where the value of the integral is $\sim i \pi/2\sqrt{s}$ with the principal
value negligible in the region of interest ($\sqrt{s} \ll M_s$) 
and we have considered constant coupling suppression $c$ 
for the modes with $n\geq 2$ (approximation that turns out to be
reasonable as the coupling saturates  quickly as we consider
higher and higher levels).
The first state is singled out because of
its different coupling.

\subsection{$e^+e^-\rightarrow \gamma +
\mbox{\textit{missing energy}}$  process}

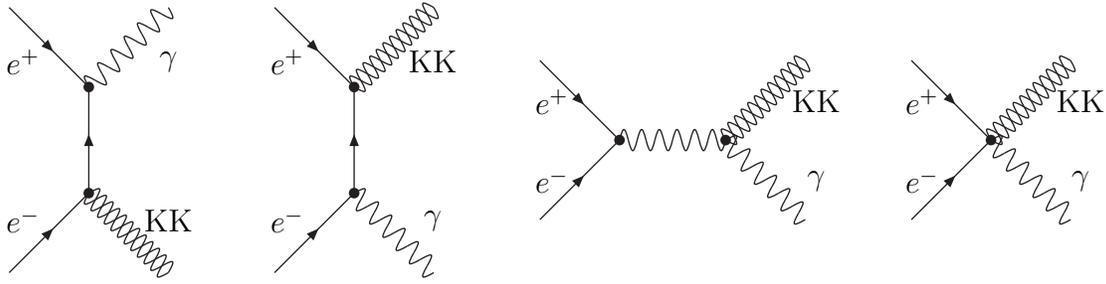
\begin{figure}[t]
\begin{center}
\begin{picture}(300,100)(0,50)

\ArrowLine(-55,50)(-25,80)
\ArrowLine(-55,150)(-25,120)
\ArrowLine(-25,80)(-25,120)
\Vertex(-25,80){2}
\Vertex(-25,120){2}
\Photon(-25,120)(5,150){4}{6}
\Photon(-25,80)(5,50){4}{6}
\Photon(-25,80)(5,50){-4}{6}
\Text(-50,130)[]{$e^+$}
\Text(-50,70)[]{$e^-$}
\Text(5,130)[]{$\gamma$}
\Text(5,70)[]{KK}

\ArrowLine(45,50)(75,80)
\ArrowLine(45,150)(75,120)
\ArrowLine(75,80)(75,120)
\Vertex(75,80){2}
\Vertex(75,120){2}
\Photon(75,80)(105,50){4}{6}
\Photon(75,120)(105,150){4}{6}
\Photon(75,120)(105,150){-4}{6}
\Text(50,130)[]{$e^+$}
\Text(50,70)[]{$e^-$}
\Text(105,130)[]{KK}
\Text(105,70)[]{$\gamma$}

\ArrowLine(145,70)(175,100)
\ArrowLine(145,130)(175,100)
\Vertex(175,100){2}
\Photon(175,100)(215,100){4}{6}
\Vertex(215,100){2}
\Photon(215,100)(245,70){4}{6}
\Photon(215,100)(245,130){4}{6}
\Photon(215,100)(245,130){-4}{6}
\Text(150,115)[]{$e^+$}
\Text(150,85)[]{$e^-$}
\Text(250,115)[]{KK}
\Text(250,85)[]{$\gamma$}

\ArrowLine(285,70)(315,100)
\ArrowLine(285,130)(315,100)
\Vertex(315,100){2}
\Photon(315,100)(345,70){4}{6}
\Photon(315,100)(345,130){4}{6}
\Photon(315,100)(345,130){-4}{6}
\Text(290,115)[]{$e^+$}
\Text(290,85)[]{$e^-$}
\Text(350,115)[]{KK}
\Text(350,85)[]{$\gamma$}

\end{picture}
\caption{$e^+e^-\rightarrow \gamma$  KK}\label{miss}
\end{center}
\end{figure}

The missing energy processes in the SM (\textit{i.e.}
$e^+e^-\rightarrow \gamma\nu\bar{\nu}$) are well explored and are a
standard way to count the number of neutrino species. In the presence
of the KK modes there is also a possibility that any KK mode produced, if it has large enough
lifetime, escapes from the detector before decaying, thus giving
us an additional missing energy signal. The new diagrams that
contribute to this effect are the ones in the Fig. \ref{miss}.

The differential cross section of the production of a KK mode plus a photon is
given by Ref. \cite{missing} and is equal to:
\be
\frac{d\sigma}{dt}(e^+e^-\rightarrow \gamma +  {\rm{KK}})=\frac{\alpha}{16}\sum_{n>0}\frac{1}{c_n^2s}F\left(\frac{t}{s},\frac{m_n^2}{s}\right)
\ee
where $s$, $t$ are the usual Mandelstam variables and the function $F$ is given by:
\ba
F(x,y)=\frac{1}{x(y-1-x)}[&-4x(1-x)(1+2x+2x^2)+y(1+6x+18x^2+16x^3)\nonumber\\
&-6y^2x(1+2x)+y^3(1+4x)\phantom{0}]
\ea

A reasonable size of a detector is of the order of $d=1$m, so we assume that the events of KK production are counted as missing
energy ones if the KK modes survive at least for distance d from the
interaction point (this excludes decays in neutrino pairs which always
give missing energy signal). We can then find a limit on the KK masses that contribute
to the experimental measurement. By a straightforward relativistic
calculation we find that this is the case if:
\be
\Gamma_n< \frac{E_\gamma}{m_nd}=\frac{s-m_n^2}{2\sqrt{s}m_nd}
\ee
From equation \ref{gam} we see that this can be done if:
\be
m_n<\sqrt{\frac{-c_n^2+c_n\sqrt{c_n^2+8\beta ds^{3/2}}}{4\beta d \sqrt{s}}}
\ee
It turns out that usually only the first KK state mass satisfies this
condition and decays outside the detector. All the other states have
such short lifetimes that decay inside the detector and so are not
counted as missing energy events (again this excludes decays in neutrino pairs). In the regions of the parameter
space where this was not the case, we found that only a very small part of
the KK tower contributed and didn't give any important excess in
comparison with the one from the first state alone. 
Thus, taking only the contribution of the first KK state and imposing
 the kinematic cuts given by the  
experiments on the angular integration, we found the measurable cross section. This cross section
has to be compared to the error of the experimentally measured cross
section because so far the SM predictions coincide with the measured
value.

The most stringent measurement available is the one by OPAL
Collaboration \cite{OPAL} at $\sqrt{s}=183$GeV. The measured cross section
is $\sigma_{meas}=4.71\pm 0.34$pb so the values of the parameters of
the model that give cross section greater than $0.34$pb are
excluded. Since the main contribution comes from  the first KK state and
because its coupling depends only on the warp factor $w$, we will either
exclude or allow the whole k-x plane for a given $w$. The critical
value of $w$ that the KK production cross section equals to the
experimental error is $w=1.8e^{-35}$. 

It is worth noting that the above cross section is almost constant for
different center of mass  energies $\sqrt{s}$, so ongoing experiments with
smaller errors (provided that they are in accordance with the SM
prediction) will push the bound on $w$ further ahead.

\subsection{Cavendish experiments}

A further bound on the parameters of our model can be put from the
Cavendish experiments. The fact that gravity is Newtonian at
least down to millimeter distances implies that the corrections to
gravitational law due to the presence of the KK states must be
negligible for such distances. The gravitational potential is the
Newton law plus a Yukawa potential due to the exchange of the KK
massive particles (in the Newtonian limit):
\be
V(r)=-\frac{1}{M_{\rm Pl}^2}\frac{M_1 
M_2}{r}\left(1+\sum_{n>0}\left(\frac{M_{\rm Pl}}{c_n}\right)^2e^{-m_nr}\right)
\ee

The contribution to the above sum of the second and
higher modes is negligible compared with the one 
of the first KK
state, because they have larger masses and coupling
suppressions. Thus, the
condition for the corrections of the Newton law to be small for
millimeter scale distances is:
\be
x<\tilde{x}=15-\frac{1}{2}{\rm ln}\left(\frac{-{\rm ln}w}{kw}{\rm
GeV}\right)
\label{Cbound}
\ee

\subsection{$k-x$ plots}

\begin{figure}[!h]

\begin{center}
\epsfig{file=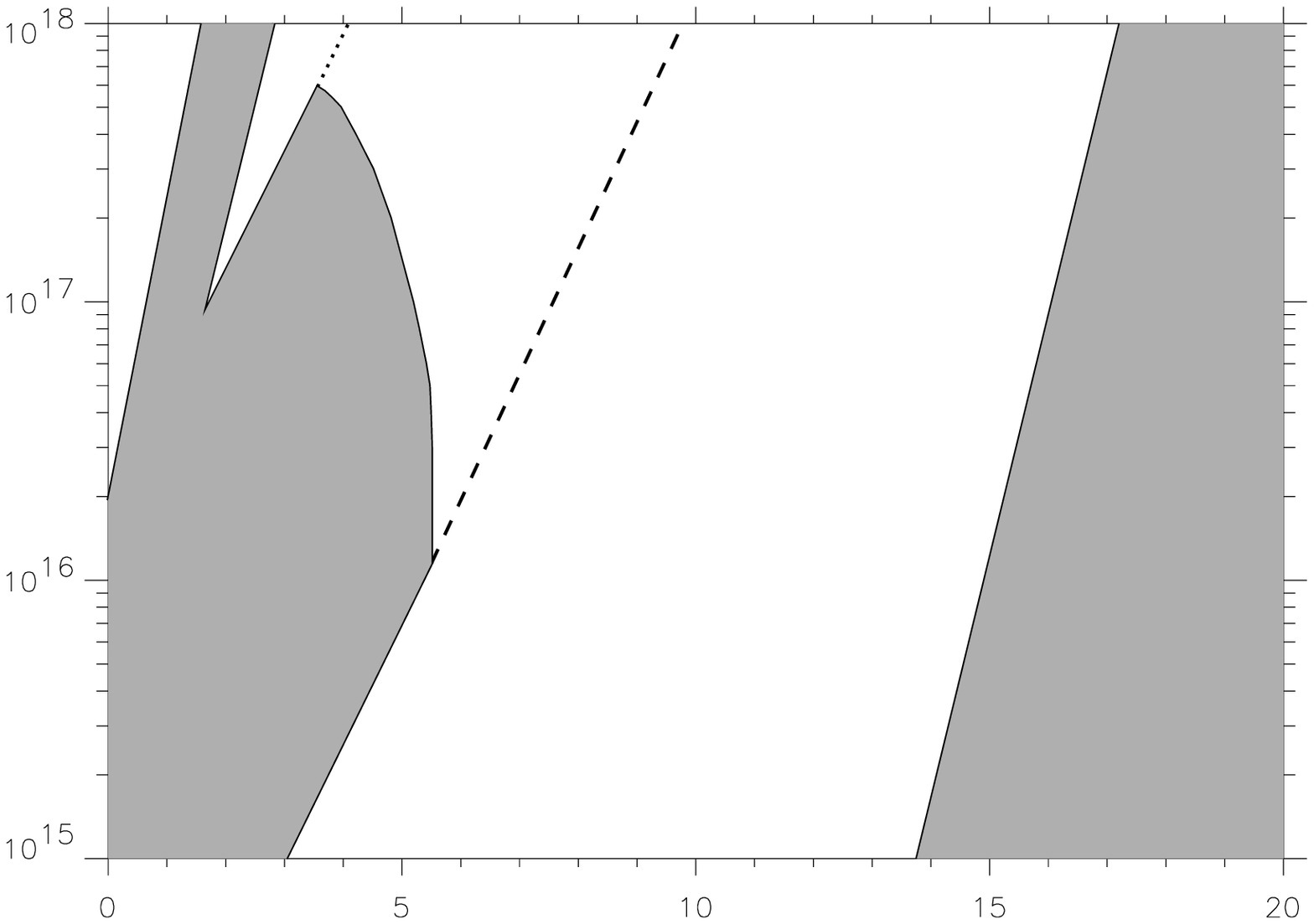,width=12cm}

\begin{picture}(-200,-200)(0,0)
\SetScale{1}

\LongArrow(-110,5)(-70,5)
\rText(-135,5)[l][]{$x$}
\LongArrow(-280,160)(-280,200)
\rText(-315,145)[l][]{$k$ (GeV)}

\rText(-250,230)[l][]{$(1)$}
\rText(-175,190)[l][]{$(2)$}
\rText(-35,145)[l][]{$(3)$}

\end{picture}

\caption{Excluded regions (shaded areas) for $w=4.5e^{-35}$.\label{plot1}}
\end{center}


\begin{center}
\epsfig{file=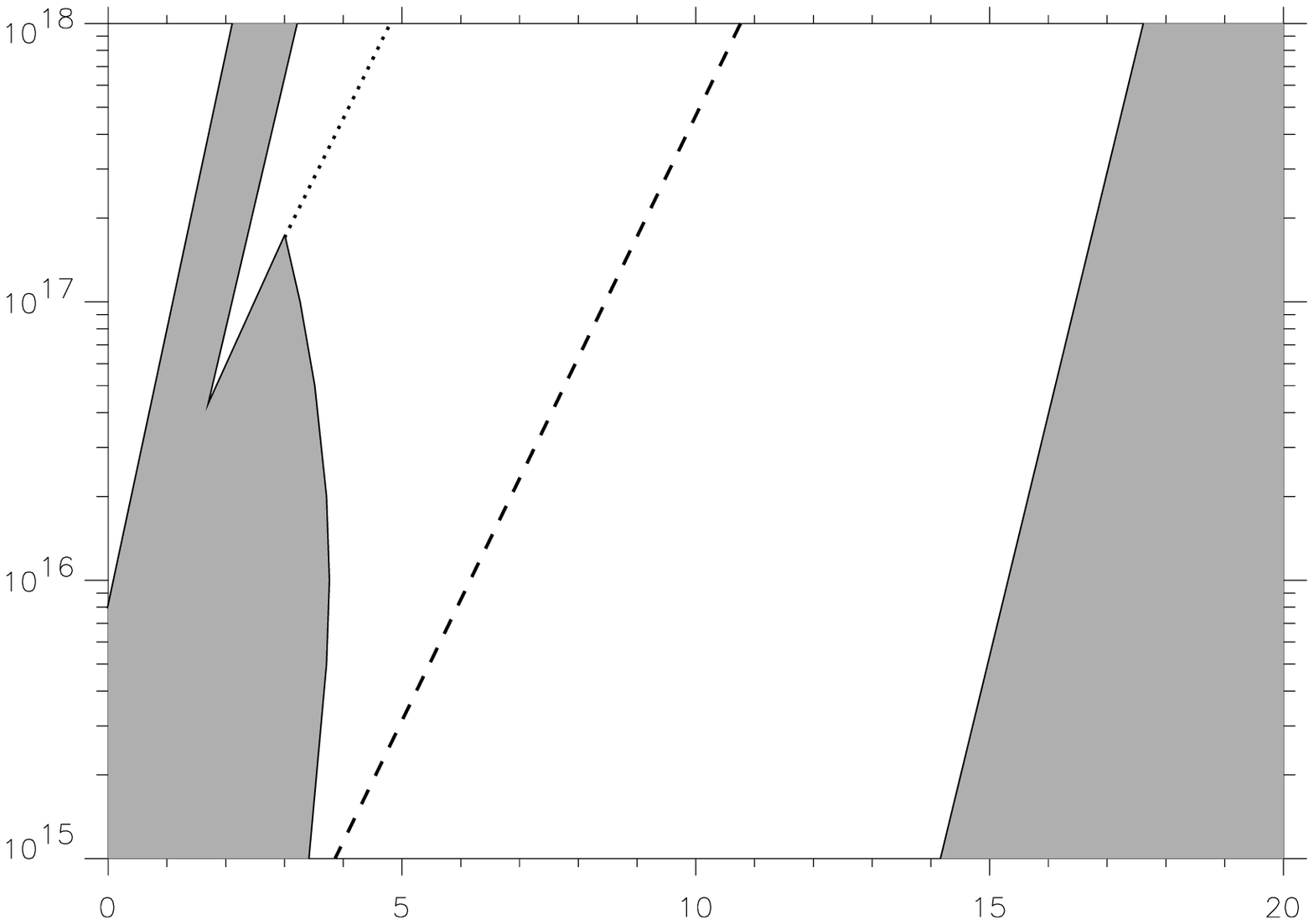,width=12cm}

\begin{picture}(-200,-200)(0,0)
\SetScale{1}

\LongArrow(-110,5)(-70,5)
\rText(-135,5)[l][]{$x$}
\LongArrow(-280,160)(-280,200)
\rText(-315,145)[l][]{$k$ (GeV)}

\rText(-245,220)[l][]{$(1)$}
\rText(-197,150)[l][]{$(2)$}
\rText(-28,145)[l][]{$(3)$}

\end{picture}

\caption{Excluded regions (shaded areas) for $w=10e^{-35}$.\label{plot2}}
\end{center}

\end{figure}

As mentioned above the  range of the parameter space that we
explore is chosen so that it corresponds to the region of physical
interest giving rise to the observed hierarchy between the electroweak
and the Planck scale
i.e. $w\sim 10^{-15}$, $k\sim M_{\rm Pl}$. The allowed  regions
(unshaded areas)
for  $w=4.5e^{-35}$ and $w=10e^{-35}$ are shown in the Figures
~\ref{plot1} and ~\ref{plot2}. The bounds from the previously mentioned
experiments and the form of the diagram will be now explained in detail.

\begin{itemize}
\item {\bf$e^+e^-\rightarrow\mu^+\mu^-$ bounds}
\end{itemize}

As we noted in section 3.1, for relatively small values of $x$ the
spectrum is discrete and as $x$ increases it tends to a continuum (the dashed line shows approximately where
we the spectrum turns from discrete to continuum).
In case of the continuum, for the parameter region that we explore, it
turns out that it does not give any bound since the excess over  the  SM cross
section becomes important only for energies much larger than $200$GeV.
However there are significant bounds coming from the discrete spectrum
region, since generally we
have KK resonances in the experimentally accessible region and  the
convolution of some of them will give significant excess to the SM
background.
The exclusion region coming from $e^+e^-\rightarrow\mu^+\mu^-$, is the
region between the curves (1) and (2).  
The details of the bound depend on the behaviour of the couplings
and the masses. In this case the bounds start when the KK states
have sufficiently large  width and height ({\it i.e.} large  mass and coupling).
This is the reason why curve (2) bends to the left as k increases.
The shape 
of the upper part of the curve (2) comes from the fact that by increasing
$k$ we push the masses of the KK states to larger values so that there
is the possibility that the
first KK state has mass smaller that $20$GeV and at the same time the 
rest of tower is above $200$GeV (the dotted line is where the second KK states is at $200$GeV). The last
region is not experimentally explored at present.
An increase of $w$  decreases all the couplings  and thus this will push
the bound even more to the left.
Decreasing $w$ ({\it e.g.} $w=e^{-35}$), on the contrary will increase the values of the
couplings and there are strict bounds coming both from the discrete and
the continuum.
The $e^+e^-\rightarrow\mu^+\mu^-$ experiments don't give any bound when the first KK
state has mass bigger than $\sim200$GeV, since in this case the KK
resonances cannot be produced from current experiments and the low
energy effects are negligible for the range of parameters that we
examine (this region is represented by the triangle at the upper left
corner of the plot).  As colliders
probe higher center of mass energies the curve (1) will be pushed to
the left and curve (2) to the right.

\begin{itemize}
\item {\bf Missing energy bounds}
\end{itemize}

As we noted in section 3.2, the KK states have generally very
short lifetime. For certain value of $w$, the main contribution to
the cross section comes from the first KK state, since the restriction
on the mass (so that the KK states escape
the detector) means that only a few states contribute, even near
the boundary of Cavendish bounds where the spacing of the tower is
very small.
Decreasing the value of $w$, we increase the coupling of
the first KK state so for the values of $w<1.8e^{-35}$ the contribution
from the first KK state becomes so big that excludes all the region
between $e^+e^-\rightarrow\mu^+\mu^-$ and the Cavendish bound (the region
with the KK tower over $200$GeV always survives).
Variation of the $k$, $x$ parameters in this case does not change
significantly the cross section, because the main contribution comes from
the first KK mode  whose coupling is constant {\it i.e.} independent of $k$,
$x$ and since although its mass, $m_1$ depends on $k$ the cross section is
insensitive to the mass changes because  it is evaluated at energy
$\sqrt{s}=183$GeV where the cross section has saturated.
To summarize, the missing energy bounds either exclude the whole
region between the $e^+e^-\rightarrow\mu^+\mu^-$ and the Cavendish limits or
nothing at all (due to the smallness of the coupling).
Additionally, the missing energy experiments don't give any bound when the first KK
state has mass bigger than $\sim143$GeV, since in this case the emitted
photon has energy smaller than the experimental cuts. Currently
running and forthcoming colliders will push the bound of $w$ to larger
values.

\begin{itemize}
\item {\bf Cavendish bounds}
\end{itemize}

From the discussion in section 3.3 we see that the bound on the parameter
space from Cavendish experiments comes from Eq. (\ref{Cbound}). The exclusion
region is the one that extends to the right of the line (3) of the
plots.  
  For fixed $k$, $w$  the Cavendish bounds exclude the region $x>\tilde{x}(k,w)$ due to the fact
that the first KK becomes very light (and its coupling remains
constant). Now if we increase $k$, since the masses of the KK are
proportional to it, we will have an exclusion region of
$x>\tilde{x}(k',w)$, with $\tilde{x}(k',w)>\tilde{x}(k,w)$. This explains the form of the
 Cavendish bounds. When we increase the $w$ parameter the whole bound
will move to the right since the couplings of the KK states
decrease. Future Cavendish experiments testing the Newton's law at
smaller distances will push the curve (3) to the left.

\section{Conclusions}

In this paper we discussed the phenomenology of the $''+-+''$
model, presented in \cite{mill}.
This was done by
considering processes sensitive to the new physics, namely the $e^+e^-
\rightarrow \mu^+\mu^-$ and the
$e^+e^- \rightarrow \gamma {\rm KK}$.
We examined the parameter space that is of physical interest
({\it i.e.} creates the hierarchy between the $M_{\rm Pl}$ and $M_{\rm E/W}$). Different regions of the parameter space 
are sensitive to different experiments. The previous experiments  
in combination with the latest Cavendish experiments place bounds
on the parameter space of the model, as shown
in Figs. \ref{plot1} and \ref{plot2}. As it can be seen there is
still a lot of the parameter space that is not presently excluded.

{\bf Acknowledgments:} We are grateful to our supervisors  Ian I. Kogan
and Graham G. Ross for useful discussions and comments and to Peter
B. Renton for providing us the LEP beam spreads. S.M.'s work is
supported by the Hellenic State Scholarship Foundation (IKY) \mbox{No. 
8117781027}.
A.P.'s work is supported by the Hellenic State Scholarship Foundation
(IKY) \mbox{No. 8017711802}.

\end{document}